\documentclass[12pt,a4paper]{article}

 \usepackage[dvips]{graphics}
 \usepackage{graphicx}
 \usepackage{afterpage}
 \usepackage{enumerate}
  \usepackage{longtable}
\begin{document}

 \title{
 Investigation of the magnetic  and temperature \\ sensitivity
   of the Stokes  parameters  of absorption \\ lines in the solar photosphere }

 \author{V.A. Sheminova}
 \date{}

 \maketitle
 \thanks{}
\begin{center}
{Main Astronomical Observatory, National Academy of Sciences of
Ukraine
\\ Zabolotnoho 27, 03689 Kyiv, Ukraine\\ E-mail: shem@mao.kiev.ua}
\end{center}

 \begin{abstract}
Response functions to perturbations in the temperature, pressure, microturbulent
velocity, and magnetic intensity were calculated for the Stokes parameter profiles
of the lines Fe I 525.06, 525.02 and Fe II  614.92~nm. The procedure proposed by
Grossmann-Doerth, Larsson, and Solanki (1988) was used. We show that the depression
response functions may be used not only to determine the depths at which changes in
the physical conditions affect most effectively the absorption and emission in the
continuum and in lines, but to estimate the response of Stokes profiles as well.
The response was estimated using sensitivity indicators calculated as an integral
of the response function over all photospheric layers. An anomalous temperature
sensitivity was found for the Stokes profiles in lines with high excitation and
ionization potentials such as the lines of O I, C I, Fe II. The depression of such
lines increases rather than decreases with growing temperature. The magnetic
sensitivity of Stokes profiles depends primarily on the magnetic field conditions.
The response of V profiles is the greatest under the weak-field and
intermediate-field conditions for photospheric lines with large values of the Land\'{e }
factor, wavelength, and equivalent width. The results of calculations of
sensitivity indicators are presented for magnetic lines together with the indices
of magnetic and temperature enhancement.

\end{abstract}

\section{Introduction}
     \label{S-Introduction}

Small-scale features often called magnetic elements or magnetic tubes have a
profound impact on the structure of the solar atmosphere. The magnetic tubes are
supposed to be involved in the energy transfer to the upper chromospheric layers
and corona, to affect the solar global magnetic fields and dynamo, to change the
solar convection characteristics, etc. Much effort has been directed in the last
decades toward thorough studies of their structure (see reviews of these numerous
studies in \cite{11,14}). Small-scale structures are difficult to study, as they
are not spatially resolved by present-day instruments. Their dimensions at the
level where photospheric lines are formed are no more than 200 km, and magnetic
intensity is more than 0.1~T. Evidence for their existence and main data on them
are obtained from the Zeeman effect manifestations observed in atomic absorption
lines which form in the photospheric regions with strong magnetic fields. Just the
observed profiles of the Stokes parameters I, Q, U, and V for absorption lines
provide the most comprehensive information on the layer structure. They are a
powerful diagnostic tool in constructing models of photospheric magnetic elements.
As a rule, the Stokes V profiles are primarily used to determine the magnetic
intensity, temperature, and velocity field inside magnetic tubes; these profiles
describe the circularly polarized radiation emerging in absorption lines.
Calculations are compared with such observed parameters of the Stokes V profiles as
the maximum amplitudes and distances between them, half-widths, asymmetry of
amplitudes and areas, zero-crossing wavelength. Pairs of lines are often used to
find the ratios of maximum amplitudes and areas for the V profiles of two lines
with different sensitivity to desired parameters of magnetic tubes. In
investigations of this kind, especially in those using line pairs, it is important
to select spectral lines in such a way that photospheric magnetic tubes might be
explored at different heights and physical conditions might be determined the most
exactly. It is rather difficult to select lines properly. In this paper we propose
to do this selection using the sensitivity indicators which we derived based on
response functions for radiation polarized in a magnetic field.

The first chapter contains the results of our calculations of response functions
for the Stokes profiles of photospheric lines, and the second chapter presents the
calculations of magnetic and temperature sensitivity indicators for the Stokes
profiles.

\section{Response functions
}

When fine effects are studied in the spectral analysis of absorption line profiles,
the contribution and response functions are used, since only these functions allow
us to find reliably the depths of absorption line formation. The contribution
functions for the Stokes profiles were investigated quite comprehensively by
Grossmann-Doerth et al.  \cite{5}, Ress et al.~\cite{9}, Solanki et al. \cite{12},
and in our paper \cite{2}. The number of studies dealing with the response
functions for the Stokes parameters is not large, they still are waiting for their
thorough investigation, which may be explained by complexity of calculations. In
1977 Landi Degl'Innocenti with coauthors  \cite{7} introduced the concept of
response functions for the Stokes parameters, regarding them as a convenient tool
for the diagnostics of velocity fields and magnetic inhomogeneities in the
atmospheres where spectral lines are formed. The method of atmosphere diagnostics
with response functions was further elaborated in  \cite{8}. Grossman-Doerth et al.
\cite{5} generalized the response functions derived by Caccin et al.  \cite{4} for
photospheric lines to the case with an arbitrary magnetic field and obtained handy
response function expressions for both emission and depression in four Stokes
parameters. We used these expressions to develop an algorithm for calculating the
response functions of the Stokes parameters for photospheric absorption lines. A
detailed description of the algorithm and calculation program is available in paper
\cite{1}. Here we give only the principal expression for a response function of a
relative depression in a polarized radiation in an absorption line (it is called a
depression response function) in the matrix representation:

   \[\mathbf{RF}_\beta=\beta \mathbf{T}^{-1} \left[\frac{\delta \mathbf{F}}{\delta\beta}-
   \frac{1}{\kappa_c}
   \frac{\delta\kappa_c}{\delta\beta}(\mathbf{AR}+\mathbf{RA}^*-\mathbf{F})-
   \frac{\delta \mathbf{A}}{\delta\beta}\mathbf{\mathbf{R}}
   -\mathbf{R}\frac{\delta \mathbf{A}^*}{\delta\beta}\right](\mathbf{T}^*)^{-1}.\]
 Here $\beta$ is a photospheric parameter which experiences small disturbances
($\delta\beta/\beta\ll1$); $\mathbf{R}$ is the matrix describing a relative
depression of polarized radiation in the line (line depth),

\[\mathbf{R} = \frac{1}{2}\left|\begin{array}{cc}
RI + RQ~~~~~~~~ RU + iRV \\ RU - iRV~~~~~~~RI - RQ \\ \end{array}   \right|,\]
 where $RI = 1 - I/I_c$, $RQ = -Q/I_c$, $RU = -U/I_c$, $RV = -V/I_c$ are the Stokes
parameters in relative depression units (by analogy with the line depth, they may
be called the Stokes parameter depths). The classical Stokes parameters I, Q, U, V
represent the intensity of the polarized radiation emerging in the spectral line;
$I_c$ is the continuum intensity; $\mathbf{A}$ is the matrix describing the
polarization, absorption, and dispersion properties of the spectral line and the
medium; $\mathbf{F}$ is the source function matrix; $\mathbf{T}$ is the matrix from
the additional equation $d\mathbf{T}/d\tau = \mathbf{AT}$; $\kappa_c$ is the
continuous absorption coefficient.

Response functions for each individual Stokes profile are calculated from the
following relations for the elements of the matrix $\mathbf{RF}_\beta$:
\[ \begin{array} {l}
 RF_{\beta ,RI}=RF_{\beta ,11}+RF_{\beta ,22},\\
 RF_{\beta ,RQ}=RF_{\beta ,11}-RF_{\beta ,22},\\
 RF_{\beta ,RU}=RF_{\beta ,12}+RF_{\beta ,21},\\
 RF_{\beta ,RV}=(RF_{\beta ,12}-RF_{\beta ,21})(-i) .
 \end{array}\]
 \noindent
 Generally the depression response functions characterize an additional contribution
from each atmospheric layer to the line depression. This contribution is associated
with those changes in absorption and emission in the line or in the continuum which
result from a disturbance in one of atmospheric parameters. The depression response
functions for the Stokes profiles describe an additional depression (response) in
the polarized radiation in the given absorption line. Response functions depend
on the kind of disturbance, line profile section, and height of the atmospheric
layer. It should be noted also that the approximate equality

\[ \mathbf{RF}_\beta (\Delta\lambda, x) \approx
\Delta\mathbf{R}(\Delta\lambda,x)/(\Delta\beta/\beta(x))\]
 is true for a response function in a thin layer at the height $x = \log \tau_5$.

The response function may be used to estimate variations in the relative depression
of the emerging polarized radiation in the line ($\Delta\mathbf{R}$) caused by the
disturbance $\Delta\beta/\beta$ when the disturbance is substantially less than unity:

\[\Delta\mathbf{R}(0) =  \int \mathbf{RF}_\beta  \frac{\Delta\beta}{\beta}d\tau .\]

Numerical calculations of the response functions for the Stokes parameters were
made with the SPANSATM program  \cite{1} for the HOLMU model photosphere with a
magnetic intensity $H=0.2$~T, an angle of inclination $\gamma =30^\circ$, an
azimuth $\phi= 30^\circ$, and $\xi_{\rm mic}= 0.8$~km/s, $\gamma_{\rm VdW} =
1.3\gamma_6$. The macroturbulent velocity does not appear in the calculations of
the absorption coefficient, and we neglected it in our response function
calculations. Spectral lines were selected so that their response to magnetic field
effects might be different as much as possible. The parameters of these lines are
given in Table~1. Central depths $R$ and equivalent widths $W$ are taken from
observations of the quiet Sun. The iron abundance is 7.64.

%
{\footnotesize
 \begin{table}[t] \centering

 \parbox[b]{8cm}{

 {Table 1. Lines used in the analysis} \label{T1}

\vspace{0.3cm}}
 \footnotesize
\begin{tabular}{lcccccc}
 \hline
 Element&$\lambda $,\,nm & $\chi_e$,\,eV & $\log gf$ &$R$& $W,$\,pm &$g_{\rm eff}$
 \\
 \hline
Fe I &  525.02 &  0.12 & -4.89 &   0.71 &     6.5 &   3.0  \\
Fe I &  525.06 &  2.20 & -2.06 &   0.79 &     10.3&    1.5 \\
Fe II&  614.92 &  3.89 & -2.85 &   0.34 &     4.0 &    1.3 \\
  \hline
 \end{tabular}
 \end{table}}
 \noindent

\begin{figure}
   \centering
   \includegraphics[width=13 cm]{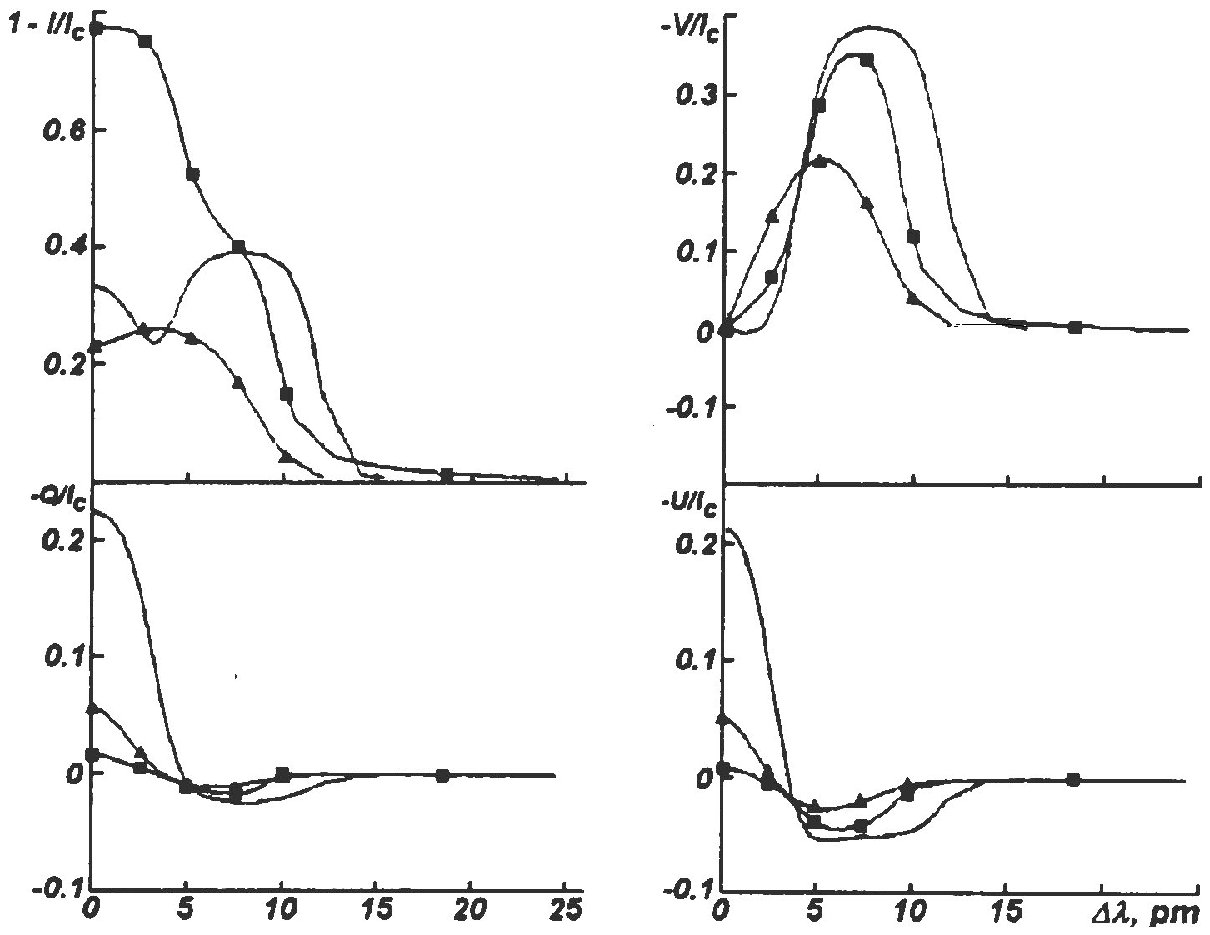}
\parbox[b]{15.5cm}{ \vspace{0.5cm}
{Fig. 1. The Stokes profiles for the lines Fe I  525.02 nm (lines), 525.06 nm
(lines and squares), and Fe II 614.92~nm (lines and triangles) calculated with the
HOLMU model solar photosphere and a magnetic field of intensity $H = 0.2$~T,
$\gamma =30^\circ$, $\phi= 30^\circ$.
 } \label{Fig1}
 }

\end{figure}

Figure 1 shows the difference between the Stokes profiles calculated for the lines
for which response functions will be found. Profiles of the V parameter are most
often used in the spectral analysis of the Stokes profiles in observations as well
as in modeling magnetic elements. Therefore we center our attention on the
properties of response functions of V profiles in our analysis of the response to
changes in the temperature, pressure, microturbulent velocity, and magnetic
intensity. Figures 2 and 3 show the most representative response functions selected
from the results of calculations. The body of data being too great, we restrict
ourselves only to the response functions for the profile section lying at the
distance $\Delta\lambda$ where the V profile attains its maximum, i.e., for
$\Delta\lambda (RV_{\rm max})$.

\begin{figure}[t]
   \centering
   \includegraphics[width=11 cm]{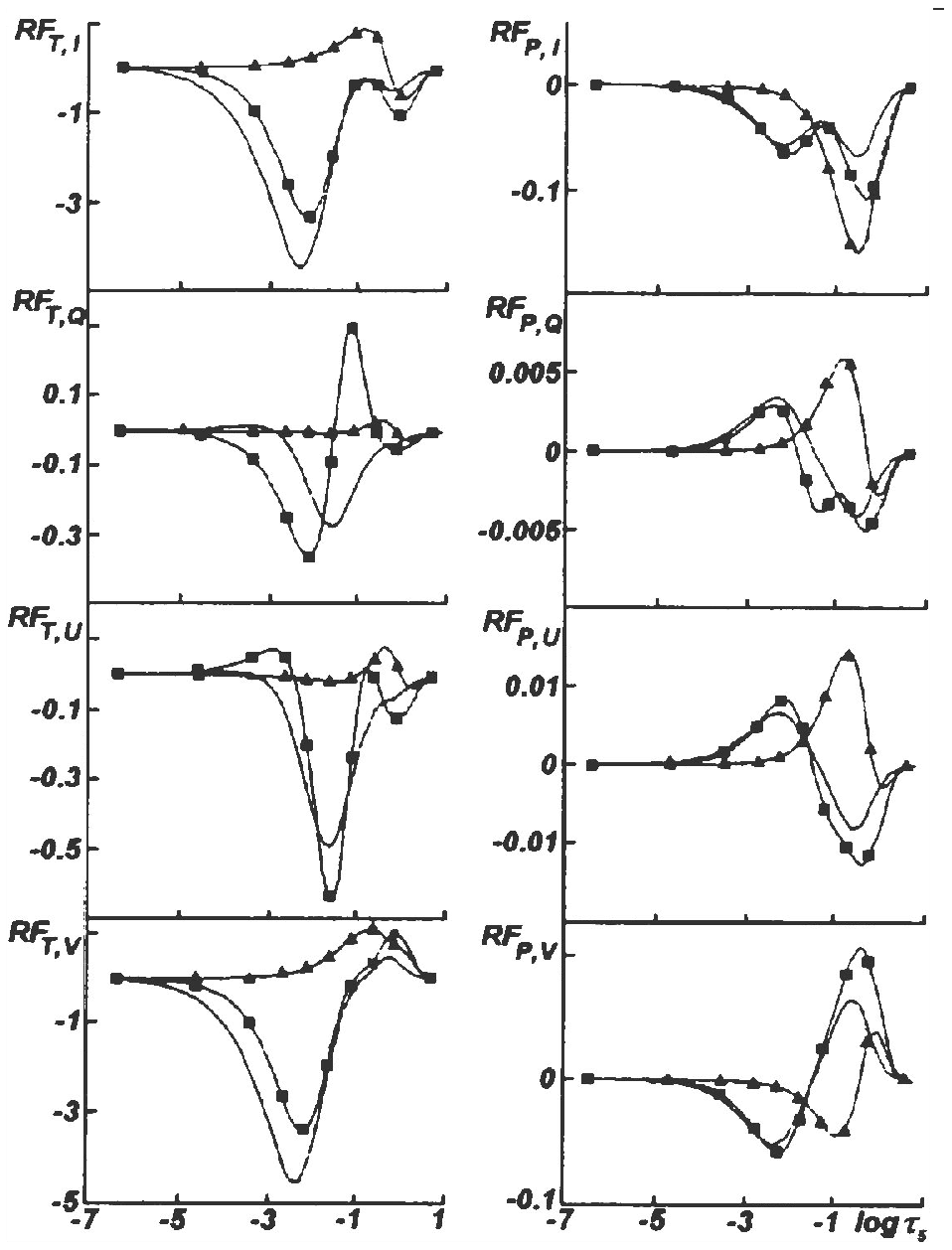}
\parbox[b]{15cm}{ \vspace{0.5cm}
{Fig. 2. Response functions for the Stokes profiles shown in Fig. 1, for the
profile section at the distance $\Delta\lambda = \Delta\lambda(RV_{\rm max})$. On
the left) response to temperature, on the right) response to pressure.  Marking is
the same as in Fig. 1.

 } \label{Fig2}
 }

\end{figure}

\subsection{Response functions to temperature variations
}

Figure 2, on the lower left, depicts the response of V profiles to temperature
variations. The magnitude of response changes from layer to layer as evidenced by
the shape of the curve describing the response function. The calculations suggest
that the shape of the curve may change from a simple one (nearly Gaussian) to a
complex one with two maxima of different heights. The greater maximum characterizes
the contribution to the V profile depression of an additional depression produced
by radiative processes in the line, while the smaller maximum at $ \tau \approx  1$
is due to radiative processes in the continuum. Comparing response functions for
different lines, one can note that the height of the greater maximum or the
integral value of the function are distinctly different. They may take both
positive and negative signs. The negative value of the response function means that
the depression $RV_{\rm max}$ decreases when the temperature grows by $\Delta T$,
and the positive value means that  $RV_{\rm max}$ increases. The integral value of
the functions, which determines the quantity  $\Delta RV(0)$, is always negative
for lines with a very small effective excitation potential $\chi_*$ (low excitation
potential plus ionization potential), it approaches zero with growing $\chi_*$ and
then increases again and becomes positive, remaining small. In practice this
suggests that in lines with high $\chi_*$ (the lines of iron ions and carbon) the
observed values of $RV$ and $RI$ may somewhat grow rather than decrease in the
regions with an increased temperature (plages), while they may become smaller in
regions with a decreased temperature (sunspots). Certainly, this reverse effect is
small, and these lines have a low sensitivity to temperature. Nevertheless this
peculiarity allows the efficiency of V profile observations to be higher when lines
with anomalous temperature sensitivity are selected.

The response functions for I profiles (Fig. 2, on the upper left) differ from the V
profile functions discussed above by the sign of the smaller maximum. This may
reflect on the integral sensitivity of I and V profiles. The general additional
negative depression in I profiles will be greater than in V profiles while the
positive depression will be smaller. The functions for Q and I profiles have more
complex shapes, and the maxima in them are smaller. In their general outline, the
response functions for I profiles are nevertheless closer to the V-profile
functions. When the magnetic vector inclination is small (it is $30^\circ$ in our
specific example), $RQ$ and $RU$ are small in magnitude at the distance
$\Delta\lambda (RV_{\rm max})$ from the line center, and therefore their
contribution to the general depression $RI$ is small. Thus one can well understand
why the function for $RI$, which represents the general depression response,
resembles the function for V profiles.

\subsection{Response functions to variations in pressure
}

The response of Stokes profiles to changes in pressure is much weaker than the
response to temperature variations. The functions of response to pressure
fluctuations are plotted at the right of Fig. 2. In these functions for V profiles
the additional depression due to processes in the continuum is greater in magnitude
than the depression due to processes in the line, i.e., the second maximum is
greater in deep layers than the first one. Interestingly, there are essential
differences between the response functions for I and V profiles. For instance, the
integral response for the ion line is almost zero in the V profile, while in the I
profile it is the greatest as compared to the other lines. As the pressure
increases, the greatest response in V profiles is found in broad lines rather than
in lines with large $\chi_*$.

\subsection{Response functions to variations in the microturbulent \\ velocity
}

A distinguishing feature of these functions is their less intricate shape (Fig. 3,
at the right). They have a well-defined maximum. This is a consequence of the fact
that disturbances in the velocity field parameters and in the magnetic field do not
affect the processes in the continuum, and therefore there is no maximum in the
deep layers. The response functions for I and V profiles are similar. The amplitude
of V profiles may diminish with increasing $\xi_{\rm mic}$ when the line half-width
is less than $\Delta\lambda_H$, and it may grow when the half-width is greater than
$\Delta\lambda_H$. The integral response is the greatest in lines with a half-width
greater than $\Delta\lambda_H$. The maximum response to $\xi_{\rm mic}$ is nearly
to the maximum response to $P_g$.

\begin{figure}[t]
   \centering
   \includegraphics[width=11 cm]{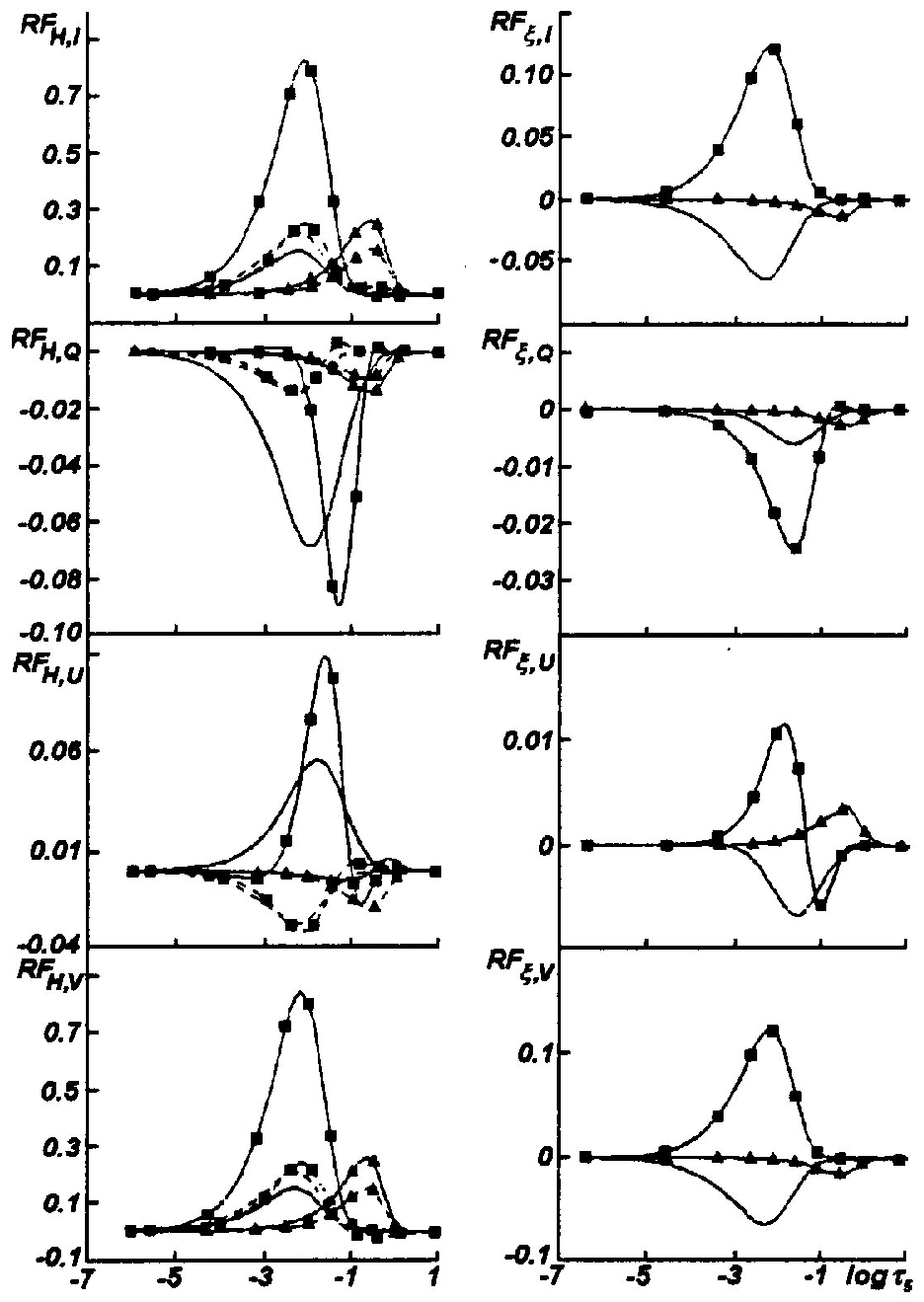}
\parbox[b]{15cm}{ \vspace{0.0cm}
{Fig. 3. The same as in Fig. 2. On the left) response to magnetic intensity, on the
right) response to microturbulent velocity. Dashed lines are depression
contribution functions.

 } \label{Fig3}
}

\end{figure}

\subsection{Response functions to variations in the magnetic intensity
}

These functions resemble in their shape the response functions for velocity
variations and differ sometimes in their sign (Fig. 3, at the left). The response
functions for I and V profiles are also similar. They are only positive, which
testifies that the depression in I and  V profiles grows on the section
corresponding to the maximum value of $RV$ when the magnetic intensity $H$
increases by $\Delta H$. The greatest response is found in the broad line with the
half-width $\geq  \Delta\lambda_H$. The maximum response to changes in the magnetic
intensity is smaller in magnitude than the response to temperature and is slightly
greater than the response to pressure and microturbulence.

\subsection{Discussion on the response functions \\ for the Stokes parameters
 }

Thus the depression response functions for the Stokes parameters of iron lines
allow us to analyze the sensitivity of profiles to structural inhomogeneities.
Among the atmospheric parameters such as the temperature, pressure, microturbulent
velocity, and magnetic intensity which govern the Stokes profiles, it is precisely
the temperature disturbances that produce the strongest response. The response to
temperature variations estimated by either the function's maximum value or its
integral value may exceed the response to magnetic intensity disturbance by a
factor about 3 and the response to pressure and velocity by a factor about 10.
Hence it follows that the temperature sensitivity should be checked first when we
determine the magnetic sensitivity of Stokes parameters, and only afterwards, if
the line is broad, the sensitivity to velocities and pressure should be checked.

The dependence of the temperature sensitivity of absorption line profiles on the
low excitation potential and ionization potential which was shown to exist in
\cite{3} is confirmed for the Stokes profiles as well. The most sensitive to
temperature are the V profiles of lines with low values of the excitation and
ionization potentials. Their amplitude diminishes with growing temperature. The
temperature sensitivity of the V profiles for lines with high values of the
excitation potential diminishes, becomes zero, and may even become anomalous, i.e.,
it grows again but with the opposite sign for very high potentials. The V profile
amplitude grows in this case. An increase in temperature in the region of
absorption line formation diminishes the maximum amplitude of V profiles for the
lines of Fe I, Ti I, Co I, Sc I, V I, Cr I, Cu I, Ni I, leaves it virtually
unchanged for the Si I, Zn I lines, and increases it slightly for the lines of C I,
0 I, Fe II, etc. The magnetic sensitivity of the Stokes V profiles depends largely
on the ratio between the line half-width and the width of the Zeeman splitting (the
same is true for the sensitivity to pressure and velocities). When the half-width
is close to or greater than the Zeeman splitting, the maximum and the area of the
response function increase abruptly, and this means that the magnetic sensitivity
increases. Hence it follows that the line magnetic sensitivity depends on the
magnetic field conditions and it cannot be forecast accurately based only on the
atomic parameters of the line.

The depression response functions are useful also in finding the depths at which
the action of the magnetic field on different sections in the Stokes profiles of
spectral lines is the most effective. The dashed lines in Fig. 3 show for
comparison the depression response functions for I and V profiles which are
recommended in  \cite{2} to be used for calculating the depths of formation of the Stokes
profiles. These functions are seen to point to the same depth region in the
photosphere. However, the calculation of response functions being substantially
more complicated, the contribution functions are likely to be more adequate for
determining the depths of formation.

Thus, the response functions, when used in the spectral analysis of the radiation
polarized in a magnetic field, provide a way for estimating the response of the
Stokes profiles to disturbances of physical conditions in a medium in all layers in
the region where the line is formed; the depths in the atmosphere at which the
disturbance changes the profiles most strongly can be appreciated also. To
determine the sensitivity of the profiles of magnetic lines, we may use estimates
of the response of Stokes profiles to changes in atmospheric parameters. We discuss
this problem in the following section.

\section{Sensitivity indicators and  indices
of magnetic \\ and temperature enhancement }

In order to determine, for example, the magnetic
sensitivity of a line, which depends not only on the Land\'{e} factor but on the
magnetic saturation of the line and its temperature sensitivity as well, and this
was clearly demonstrated in  \cite{13} by the example of the Stokes V  profiles observed in
different lines, we need to know all responses to changes in the physical
conditions. We have a possibility to analyze quantitatively the magnetic
sensitivity of Stokes parameters using the response functions which allow us to
study separately the effect of physical parameters on the amplitude, area, any
section of the Stokes profile. The integral response function, which yields the net
response to a disturbance of all photospheric layers where the absorption line is
formed, may serve as quantitative measure of the magnetic sensitivity; we call this
measure the sensitivity index or indicator for the Stokes parameters.

For the V profile, the indicator of sensitivity to the magnetic intensity $H$ (we
denote it by $P_{H,V}$) on the profile section $\Delta\lambda$ is equal to the
integral of the corresponding depression response function $RF_{H,V}$ for the
Stokes parameter V over all layers $x = \log \tau_5$:

\[ P_{H,V}(\Delta\lambda) = \int_{-\infty}^\infty RF_{H,V}(x, \Delta\lambda)dx.\]

The sensitivity index calculated in this way is the rate of variation in the
depression $RV(0,\Delta\lambda)$ emerging at the surface as related to the relative
rate of variation in the disturbed atmospheric parameter H. When we want to obtain
the index in the relative depression units, $P_{H,V}$ should be divided by the
relative depression $RV = -V/I_c$ of the Stokes parameter V. The sensitivity
indices for other Stokes parameters are determined in the same way. The sensitivity
indicator for the entire Stokes profile as a whole

\[ PW_{H,V} = \int_{-\infty}^\infty RFW_{H,V}(x)dx \]
 is calculated using the integral response function

\[ RFW_{H,V}(x) = \int_{\rm line} RF_{H,V}(x, \Delta\lambda)d(\Delta\lambda)\]
 describing the net response of the whole profile. The increase in the equivalent
width of a line due to magnetic field is called the magnetic line enhancement, and
so the magnetic sensitivity indicator for the equivalent width of the Stokes
parameter I specified by the integral

\[ PW_{H,I} = \int_{-\infty}^\infty RFW_{H,I}(x)dx \]
 is the index of the magnetic line enhancement. Recall that the approximate equality
$\Delta W_I \approx PW_{H,I}(\Delta H/H)$ is valid when disturbances are
sufficiently small $(\Delta H/H <\ll 1)$, and it allows estimating the change in
the magnetic field $\Delta H/H$ in the region of line formation if $\Delta W_I$, is
known from observations.

In calculating specific sensitivity indicators we used the same initial data as in
the calculations of response functions. The calculations of sensitivity indicators
for actual lines yielded results so varied that we could see no regularities at
first glance. We had to change the calculation tactics. First we restricted our
consideration to the sensitivity parameters for the V profiles and for the
temperature and magnetic intensity only. The profiles Q and U are observed less
accurately, as a rule, the ratio of their amplitudes to noise distortions being too
small, and they are not practically used in the Stokes profile interpretations.
Second, to find the dependence of the indicators on line atomic parameters and on
conditions of line formation, we selected hypothetical lines of Fe I with the
following parameters: $ W = 2$, 4, 8, 10, 12, 14 pm; $\lambda = 450$, 600, 750 nm;
excitation potentials $\chi_e = 0$, 2, 4 eV; the Land\'{e} factors $g_{\rm eff} =
1$, 2, 3. The reference line had the parameters $W = 4$ pm, $\lambda = 600$ nm,
$\chi_e = 2$ eV, $g_{\rm eff}= 2$, $H = 0.2$~T, $\gamma = 30^\circ$, $\phi =
30^\circ$. Varying $\lambda$, $g_{\rm eff}$, $\chi_e$, $W$, and $H$ one after
another, we calculated the V profiles and their magnetic and temperature
sensitivity. Figures 4--7 show the principal results.

\begin{figure}
   \centering
   \includegraphics[width=13 cm]{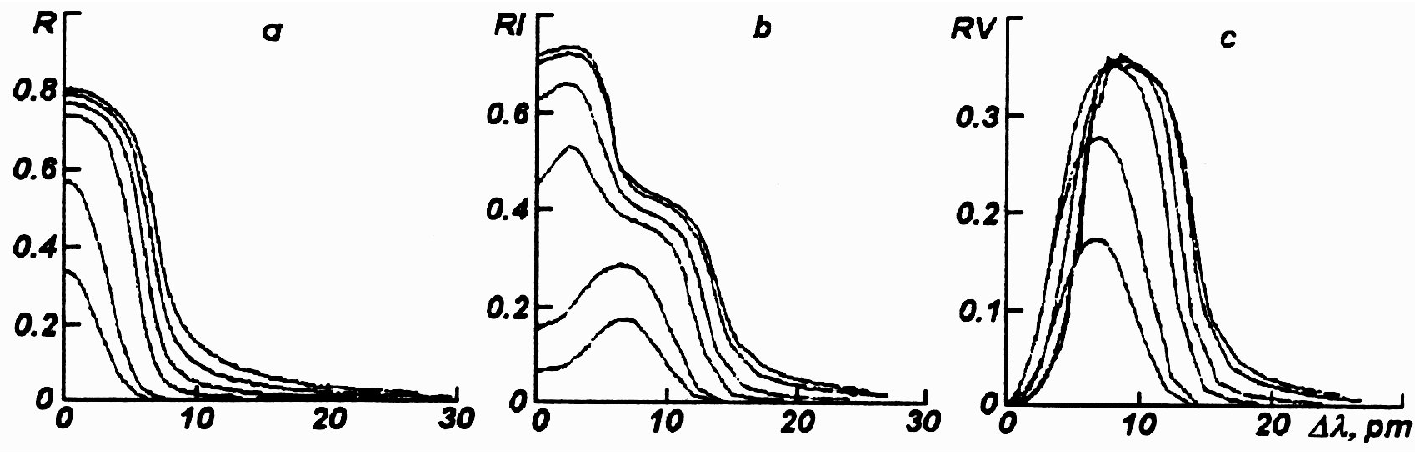}
\parbox[b]{15cm}{ \vspace{0.5cm}
{Fig. 4. Profiles of the hypothetical lines Fe I 600.0 nm, $\chi_e = 2$~eV, $g_{\rm
eff}= 2$ with $W = 2$, 4, 8, 10, 12, 14~pm calculated without a magnetic field (a)
and the Stokes profiles $RI$ (b) and $RV$ (c) in the magnetic field with an
intensity of 0.2~T, $\gamma=  30^\circ$, $\phi = 30^\circ$. HOLMU model
photosphere.

 } \label{Fig4}
 }

\end{figure}
\begin{figure}
   \centering
   \includegraphics[width=13 cm]{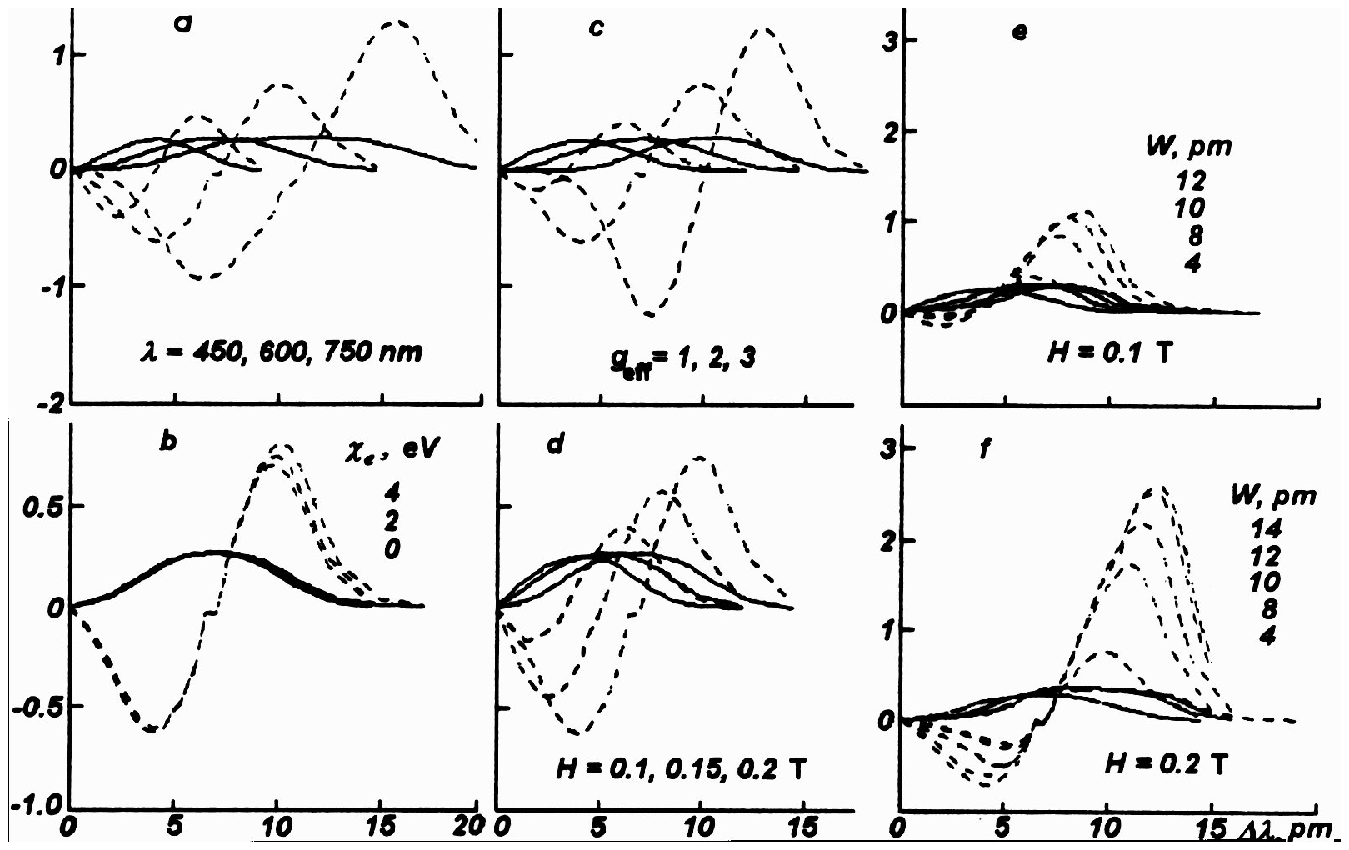}
\parbox[b]{16cm}{ \vspace{0.5cm}
{Fig. 5. Magnetic sensitivity profiles (dashed lines) calculated for the Stokes V
profiles (solid lines) in the magnetic field with $H = 0.2$~T, $\gamma = 30^\circ$,
$\phi = 30^\circ$: a) with different wavelengths and $ \chi_e = 2$~eV, $g_{\rm eff}
= 2$, $W = 3$, 4.4, 6.6 pm; b) with different excitation potentials and $\lambda =
600$~nm, $ \chi_e=2$, $  W = 4.4$ pm; c) with different Land\'{e} factors and
$\lambda = 600$~nm, $ \chi_e=2$, $W = 4.4$ pm; d) with different magnetic
intensities and $\lambda = 600$~nm, $ \chi_e=2$, $g_{\rm eff} = 2$, $W = 4.4$ pm;
e) with different equivalent widths and $\lambda = 600$~nm, $ \chi_e=2$, $g_{\rm
eff} = 2$, $H= 0.1$~T; f) the same as (e) but with $H = 0.2$~T.

 } \label{Fig5}
 }

\end{figure}

\section{Results and discussion
}

Figure 4a shows the profiles of the hypothetical lines with different equivalent
widths. The profiles were calculated for the quiet Sun, i.e., without magnetic
field. Figures 4b,c show the profiles of the Stokes parameters I and V only, in
relative depression units, for the same lines ($RI =  1 - I/I_c$, $RV = -V/I_c$).
Similar to an absorption line profile, the sensitivity profile $P_{H,V}
(\Delta\lambda)$ can be calculated, it represents the dependence of sensitivity
indicators on the distance $\Delta\lambda$ to the line center. The sensitivity
profiles can be used for the analysis of the magnetic sensitivity of the Stokes V
profiles. Figure 5 depicts V profiles (solid lines) and magnetic sensitivity
profiles (dashed lines) as functions of the wavelength (a), low excitation
potential (b), Land\'{e} factor (c) magnetic intensity (d), and equivalent width in
a field with $H = 0.1$~T (e) and $H = 0.2$~T (f). The distinctive feature of the
magnetic sensitivity profile is that it is a double-peaked curve with a positive
and a negative maxima which accounts for actual variations in the Stokes V profile
when the magnetic intensity $H$ grows by $\Delta H$. The V profile given by our
calculations being antisymmetric, we consider here only its red wing which, in its
turn, also has two wings. The left wing in the V profile is always matched by a
negative peak in the sensitivity profile, and this means that $RV$ decreases due to
its shift resulting from an increase in $H$. The right wing is matched by a
positive peak which characterizes the increase in $RV$ in these line profile
sections. The distance between the peaks characterizes the width of the red wing in
the V profile. If lines have different excitation potentials and all other their
parameters are close, the magnetic sensitivity profiles of these lines are
virtually the same. The most interesting situation was observed for lines with
different equivalent widths. For a specific magnetic field, 0.2~T in this case
(Fig. 5f), the positive peak in the sensitivity profile rapidly increases and
becomes dominant when W increases and other line parameters, apart from $gf$,
remain unchanged. At the same time the negative peak grows first until $W = 8$ pm
and then it decreases and becomes much smaller than in weak lines. As the integral
sensitivity rises sharply over the entire profile, $RV_{\rm max}$ also increases
together with the V-profile area and the index of magnetic line enhancement. This
increase stops in lines with $W > 12$~pm. Simple calculations reveal that the
growth of magnetic sensitivity depends not so much on the equivalent width as on
the radio between the line half-width (we denote it by $\Delta\lambda_D$) and the
magnetic broadening ($\Delta\lambda_H$). The sensitivity of V profile grows
abruptly when $\Delta\lambda_D \geq  \Delta\lambda_H$. The ratio of these
quantities is known to specify the magnetic field conditions under which the
spectral line exists. The sensitivity indicators calculated by us suggest that the
magnetic sensitivity of V profiles shows up in different ways depending on the
field conditions. Solanki \cite{10} describes the characteristic features of these
field conditions with the magnetic sensitivity taken into account.

1) $\Delta\lambda_D \gg \Delta\lambda_H$, weak-field conditions. In this case the
distance between the red and blue maxima in the V profile, $\Delta\lambda(V_{\rm
max})$, depends on $\Delta\lambda_D$ only. The maximum amplitude is proportional to
the magnetic intensity,  $V_{\rm max}\approx H$ and $Q_{\rm max}\approx H^2$.
Spectral lines are not splitted under these condition and I profiles provide
information on the average field ${<}H{>}$ only. These lines can be used for
measuring $H$ only in the case when the field is spatially resolved or when there
is a possibility to observe V profiles. The magnetic sensitivity of V profiles is
high.

2)  $\Delta\lambda_D \approx \Delta\lambda_H$, intermediate-field conditions. Here
$\Delta\lambda(V_{\rm max})$ and $\Delta\lambda(Q_{\rm max})$ depend on both
$\Delta\lambda_H$ and $\Delta\lambda_D$. The dependence of the maximum amplitudes
$V_{\rm max}$ and $Q_{\rm max}$  on $H$ is weaker than under the weak-field
conditions. Lines are partially split, and $H$ is measured in these conditions
using such intricate methods for the analysis of Stokes profiles as the line ratio
method, inversion methods, the Fourier transformation. The magnetic sensitivity of
V profiles may vary in these conditions from moderate to high.

3) $\Delta\lambda_D \ll \Delta\lambda_H$, strong-field conditions. Spectral lines
are completely splitted. Here $\Delta\lambda(V_{\rm max})=\Delta\lambda(Q_{\rm
max})$, $V_{\rm max}$ and $Q_{\rm max}$ are independent of $H$. The field intensity
can be obtained easily from $\Delta\lambda(V_{\rm max})$ or $\Delta\lambda(Q_{\rm
max})$. The magnetic sensitivity of V profiles is low.

\begin{figure}
   \centering
   \includegraphics[width=12.5 cm]{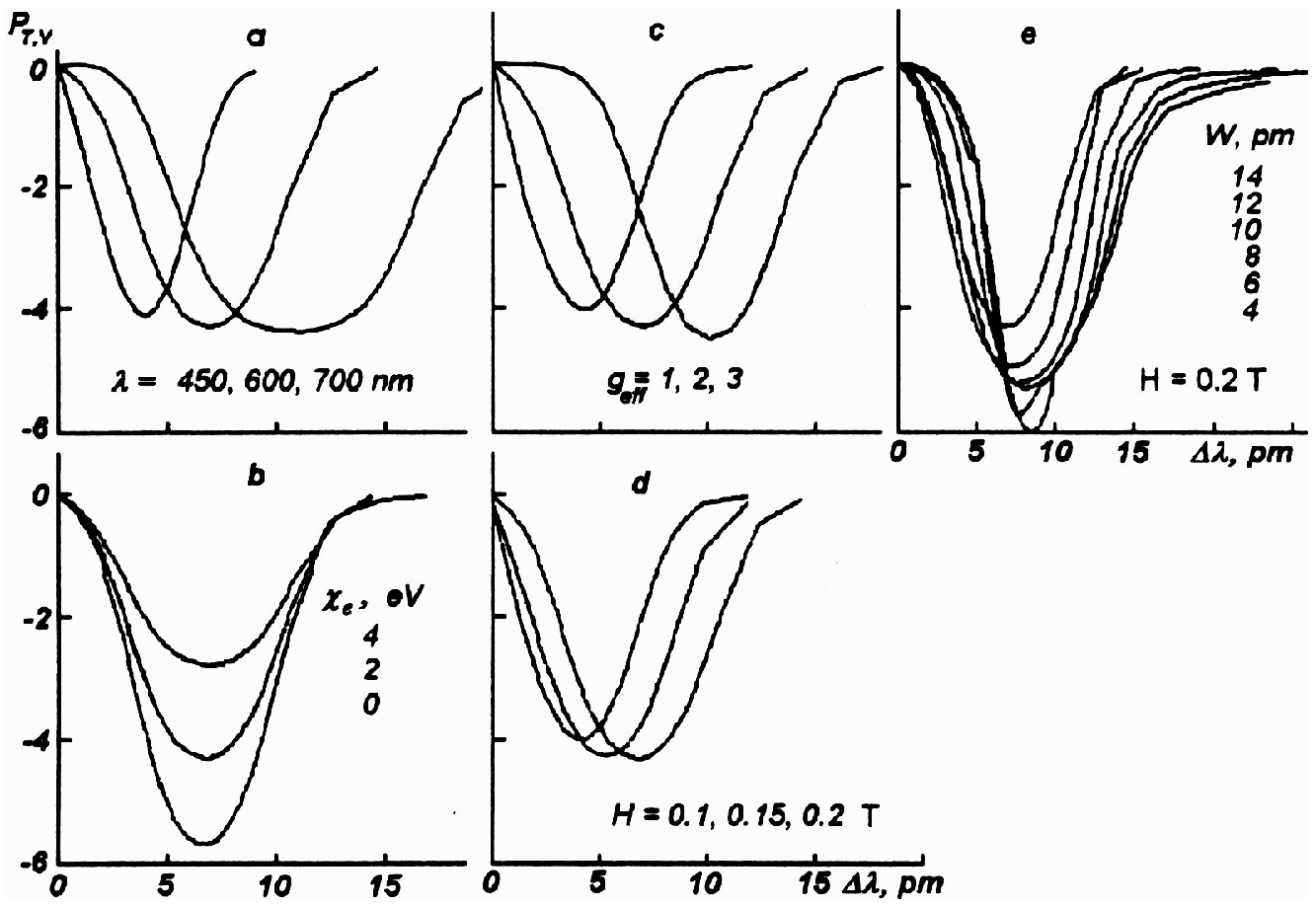}
\parbox[b]{15cm}{ \vspace{0.5cm}
{Fig. 6. Temperature sensitivity profiles calculated for the same Stokes V profiles
as shown in Fig. 5.

 } \label{Fig6}
 }

\end{figure}

All these conditions can be observed in Figs 5e,f, and the sensitivity of V
profiles for different lines can be compared. Of the lines with $W = 4$, 8, 10, 12,
14 pm which have the half-widths $\Delta\lambda_D = 3.3$, 5.2, 6.0, 6.7, 7.2 pm,
respectively, and a 6.7 pm magnetic broadening for $\lambda = 600$ nm, $g_{\rm eff}
= 2$ in the field $H = 0.2$~T (Fig. 5f), the lines with $W = 8$, 10, 12, 14 pm find
themselves under the intermediate-field conditions, while lines with $W < 8$ pm are
under the strong-field conditions. The weak-field conditions set in for lines with
$W > 14$ pm. When $\Delta\lambda_H$ becomes smaller, for instance, the magnetic
intensities reduced by one half, the magnetic sensitivity increases in lines with
much smaller equivalent widths? This means that for lines with $\lambda = 600$ nm,
$g_{\rm eff} = 2$, $H = 0.1$~T, $\Delta\lambda_H = 3.3$ pm (Fig. 5e), the lines
with $W = 4$, 6, 8 pm are in the intermediate-field conditions, lines with $W < 4$
pm are in the strong-field conditions, and lines with $W > 8$ pm are in the
weak-field conditions. In going from a strong field to a weak one, the sensitivity
of V profiles grows, reaches its peak and then remains practically unchanged.~The
same line (e.g., the line with $W=10$ pm) is under the weak-field conditions in a
field with $H = 0.1$~T and under the intermediate-field conditions at $H = 0.2$~T,
but the sensitivity of its V profile is higher where the field is stronger. The
sensitivity of lines may vary within the intermediate- and weak-field conditions
owing to differences in individual line parameters.

The sensitivity of V profiles to magnetic field may become more pronounced if lines
insensitive to temperature are selected. We have already noted in the second
section of this paper that the response of Stokes profiles to temperature
variations is by an order of magnitude greater in the most sensitive lines than the
response to magnetic intensity. The temperature strongly affects the equivalent
width $W$ and the line half-width $\Delta\lambda_D$. The latter quantity is
proportional to $\lambda T^{1/2}$, line saturation, microturbulence and damping
broadening. Changes in temperature can easily take the line from some conditions to
other and can thus change the line sensitivity.

\begin{figure}[t]
   \centering
   \includegraphics[width=13. cm]{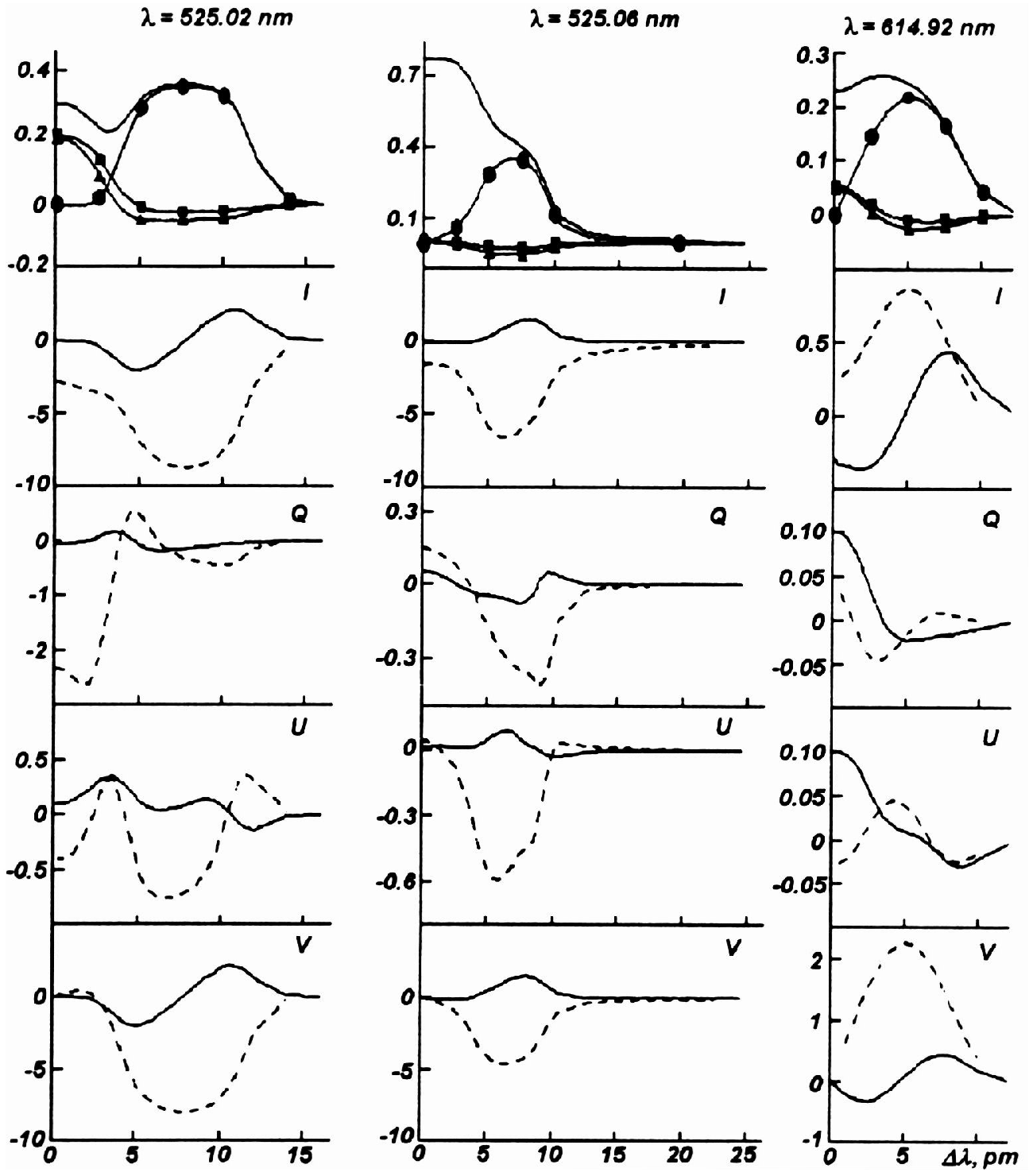}
\parbox[b]{15cm}{ \vspace{0.5cm}
{Fig. 7. Temperature (dashed lines) and magnetic (solid lines) sensitivity profiles
for the profiles of Stokes parameters for three spectral lines Fe I 525.02, 525.06
nm, Fe~II  614.92 nm in the magnetic field with $H = 0.2$~T, $\gamma = 30^\circ$,
$\phi = 30^\circ$; at the top: solid lines) $RI$, squares) $RQ$, triangles) $RU$,
dots) $RV$.

 } \label{Fig7}
 }

\end{figure}

Figure 6 presents the temperature sensitivity profiles for the V profiles of
hypothetical lines as functions of atom and medium parameters. The most
temperature-sensitive are the V profiles of moderate photospheric lines (14 pm~$ >
W > 6$~pm) with low values of excitation potential. The temperature sensitivity
profiles for actual lines are shown in Fig. 7. Besides, Table 2 gives the results
of calculations of sensitivity indicators for the maximum value of the V profile
($P_{H,V}$ and $P_{T,V}$) and for its area ($PW_{H,V}$ and $PW_{T,V}$) as well as
the indices of magnetic enhancement $PW_{H,I}$ and temperature enhancement
$PW_{T,I}$ in a field with $H = 0.2$~T, $\gamma = 30^\circ$, $\phi = 30^\circ$, and
$T$(HOLMU) for spectral lines often used in spectropolarimetric observations.~These
quantitative estimates of the temperature and magnetic sensitivity can be relied on
in a comparative analysis and in selecting lines for an investigation of the
magnetic field structure in small-scale magnetic features at the photospheric
level. To do this, we have to assess the changes in the maximum value of $RV$, in
the V profile area $W_V$ or in the line's equivalent width $W_I$, which would occur
when the magnetic intensity increases, say, by 5\% ($\Delta H = 0.01$~T, $H =
0.2$~T, $\Delta H/H = 0.05$) and the temperature increases by 2\% ($\Delta T = 100$
K, $T = 5000$ K, $\Delta T/T = 0.02$) for the lines $\lambda\lambda$ 525.02,
525.06, and 614.92 nm. We find the sensitivity indicators in Table 2 and calculate
$\Delta RV_{\rm max}$,  $\Delta W_V$, $\Delta W_I$ using the approximations of the
type $\Delta W \approx PW_H \Delta H/H$. The results are given in Table 3.

%
{\footnotesize
 \begin{table}[t] \centering
 \parbox[b]{14cm}{
{Table 2. Magnetic and temperature sensitivity indicators.
$W_V$, $W_I$, $PW_{H,V}$, $PW_{T,V}$, $PW_{H,I}$, $PW_{T,I}$ are given in pm and
 $\chi_e$ in eV.
 \label{T:2}}
\vspace{0.3cm}}
 \footnotesize
\begin{tabular}{lllllrrrrrrrr}
 \hline
 $\lambda $,\,nm & $\chi_e$ & $g_{\rm eff}$ &$RV_{\rm max}$& $W_V$ &$W_I$ &
  $P_{H,V}$ & $P_{T,V}$ &
 $PW_{H,V}$&
 $PW_{T,V}$ &
 $PW_{H,I}$ &
 $PW_{T,I}$ \\

 \hline

 477.00  C I  & 7.48 & 1.5 & 0.043 & 0.61 & 1.26 &   0.025  &   0.625  &    0.48 &    9.0    &    0.04  &     3.0           \\
 477.59  C I  & 7.49 & 2.0 & 0.053 & 0.84 & 1.47 &   0.015  &   0.769  &    0.53 &    12.6   &    0.08  &     3.7            \\
 538.00  C I  & 7.68 & 1.0 & 0.057 & 0.91 & 2.57 &   0.048  &   0.876  &    0.84 &    14.0   &    0.19  &     4.8            \\
 612.62  Ti I & 1.07 & 1.2 & 0.126 & 1.63 & 2.17 &   0.006  &   -2.210 &    0.63 &    -28.2  &    0.13  &     -40.2          \\
 612.89  Ni I & 1.68 & 1.5 & 0.155 & 1.98 & 2.45 &   -0.055 &   -0.055 &    0.47 &    -27.9  &    0.14  &     -38.3          \\
 611.16  V  I & 1.04 & 1.3 & 0.080 & 0.90 & 1.11 &   0.033  &   -1.420 &    0.14 &    -15.5  &    0.07  &     -21.2          \\
 524.75  Cr I & 0.96 & 2.5 & 0.379 & 5.84 & 9.02 &   0.470  &   -6.415 &    5.65 &    -93.2  &    4.06  &     -138.0         \\
 570.84  Si I & 4.95 & 1.5 & 0.222 & 4.26 & 9.47 &   0.269  &   -0.839 &    4.56 &    -9.6   &    3.15  &     -46.8          \\
 523.46  Fe II& 3.22 & 0.9 & 0.285 & 2.88 & 8.93 &   0.648  &   3.350  &    5.67 &    32.3   &    5.16  &     6.4            \\
 532.55  Fe II& 3.22 & 1.1 & 0.200 & 2.28 & 4.11 &   0.121  &   1.670  &    1.91 &    19.1   &    1.21  &     5.7            \\
 541.41  Fe II& 3.22 & 1.2 & 0.126 & 1.48 & 2.65 &   0.040  &   0.880  &    0.81 &    10.7   &    0.13  &     3.4            \\
 614.99  Fe II& 3.89 & 1.3 & 0.219 & 2.93 & 4.17 &   0.060  &   2.252  &    1.85 &    29.8   &    1.08  &     10.9           \\
 636.94  Fe II& 2.89 & 2.1 & 0.133 & 1.74 & 2.09 &   0.007  &   0.846  &   -0.12 &    11.3   &    -0.17 &     2.3            \\
 480.81  Fe I & 3.25 & 1.3 & 0.218 & 1.93 & 2.76 &   0.032  &   -2.700 &    1.06 &    -23.6  &    0.31  &     -37.3          \\
 523.29  Fe I & 2.93 & 1.3 & 0.331 & 3.91 & 10.34&   1.060  &   -3.610 &    9.22 &    -43.5  &    8.70  &     -79.8          \\
 524.70  Fe I & 0.09 & 2.0 & 0.369 & 5.07 & 7.31 &   0.418  &   -7.570 &    3.94 &    -101.0 &    2.86  &     -129.0         \\
 525.02  Fe I & 0.12 & 3.0 & 0.387 & 5.87 & 8.49 &   0.288  &   -7.970 &    1.45 &    -109.0 &    1.18  &     -145.0         \\
 525.06  Fe I & 2.20 & 1.5 & 0.352 & 4.42 & 11.57&   1.334  &   -4.480 &    12.30&    -56.8  &    11.60 &     -93.8          \\
 550.14  Fe I & 0.95 & 1.9 & 0.302 & 5.46 & 14.77&   0.118  &   -5.720 &    18.70&    -75.1  &    17.80 &     -125.0         \\
 550.67  Fe I & 0.99 & 2.0 & 0.370 & 6.14 & 16.64&   0.772  &   -7.230 &    20.90&    -75.5  &    19.60 &     -119.7         \\
 609.36  Fe I & 4.61 & 1.2 & 0.177 & 2.15 & 3.11 &   0.065  &   -1.630 &    1.17 &    -2.0   &    0.41  &     -37.1          \\
 609.37  Fe I & 4.65 & 1.0 & 0.124 & 1.36 & 2.03 &   0.020  &   -1.170 &    0.75 &    -12.8  &    0.14  &     -24.8          \\
 609.66  Fe I & 3.98 & 1.5 & 0.198 & 3.11 & 4.29 &   -0.025 &   -2.080 &    1.66 &    -32.1  &    0.89  &     -85.5          \\
 615.16  Fe I & 2.18 & 1.8 & 0.282 & 4.16 & 5.39 &   -0.025 &   -4.240 &    1.32 &    -61.8  &    0.86  &     -85.5          \\
 617.33  Fe I & 2.22 & 2.5 & 0.334 & 5.92 & 8.23 &   0.103  &   -4.880 &    16.70&    -80.0  &    0.96  &     -120.0         \\
 630.25  Fe I & 3.69 & 2.5 & 0.334 & 6.77 & 11.02&   0.131  &   -3.400 &    5.63 &    -61.8  &    3.78  &     -125.0         \\
 630.34  Fe I & 4.32 & 1.5 & 0.032 & 3.75 & 0.46 &   -0.016 &   -0.346 &    0.04 &    -3.9   &    -0.10 &     -7.0           \\
 643.08  Fe I & 2.18 & 1.2 & 0.314 & 4.95 & 13.29&   1.098  &   -4.270 &    13.90&    -63.3  &    13.10 &     -107.0         \\
 673.31  Fe I & 4.62 & 2.5 & 0.161 & 2.39 & 2.92 &   -0.036 &   -1.520 &    -0.21&    -22.0  &    -0.297&     -38.3          \\

  \hline
 \end{tabular}
 \end{table}}
 \noindent

The dependence of the sensitivity indicators for the Stokes profiles on various
parameters of both the line and the medium being rather intricate, the sensitivity
to magnetic field cannot be uniquely determined, i.e., we cannot find the magnetic
sensitivity of V profiles, for instance, from such line parameters as $\lambda$,
$\chi_e$, $R$, $W$, $g_{\rm eff}$. We have to know also the parameters $H$ and $T$
of the medium, i.e., the magnetic field conditions. To determine these conditions,
we have to carry out the following procedure. We calculate the line half-width and
magnetic broadening and compare them. Figure 8 depicts the plots of $\Delta\lambda
(H)$ for different $\lambda$ and $g_{\rm eff}$ and the plots of
$\Delta\lambda_D(W)$ used for the assessment of conditions for the lines of neutral
iron. The line half-width $\Delta\lambda_D$ can be estimated from these plots with
the known $\lambda$ and $W$. With the known $\lambda$, $g_{\rm eff}$ and $H$, we
can find the magnetic broadening $\Delta\lambda_H$, compare it with
$\Delta\lambda_D$, and thus determine the sensitivity of the V profile for any Fe~I
line in a medium with the $H$ and $T$(HOLMU) chosen.

%
{\footnotesize
 \begin{table} \centering
 \parbox[b]{14cm}{
{Table 3. The changes in the maximum value of $RV$, in the V profile area $W_V$,
and  in the equivalent width $W_I$, which would occur when the magnetic intensity
increases by 5\%  and the temperature increases by 2\%.
 \label{T:1}}
\vspace{0.3cm}}
 \footnotesize
\begin{tabular}{ccccccccc}
 \hline
$\lambda $,\,nm&$g_{\rm eff}$ & $\chi_e$,\,eV & $\Delta RV_H$ & $\Delta RV_T$&
 $\Delta W_{H,V}$,\,pm & $\Delta W_{T,V}$,\,pm& $\Delta W_{H,I}$,\,pm&
 $\Delta W_{T,I}$,\,pm\\

 \hline
525.02 &  3.0& 0.12&  0.014&  -0.159&    0.072 &    -2.180 &   0.059 &   -2.580 \\
525.06 &  1.5& 2.20&  0.067&  -0.090&    0.615 &    -1.136 &   0.580 &   -1.876 \\
614.92 &  1.3& 3.89&  0.003&   0.045&    0.092 &     0.596 &   0.054 &    0.218 \\

\hline
 \end{tabular}
 \end{table}}
 \noindent
\begin{figure}[t]
   \centering
   \includegraphics[width=9.8cm]{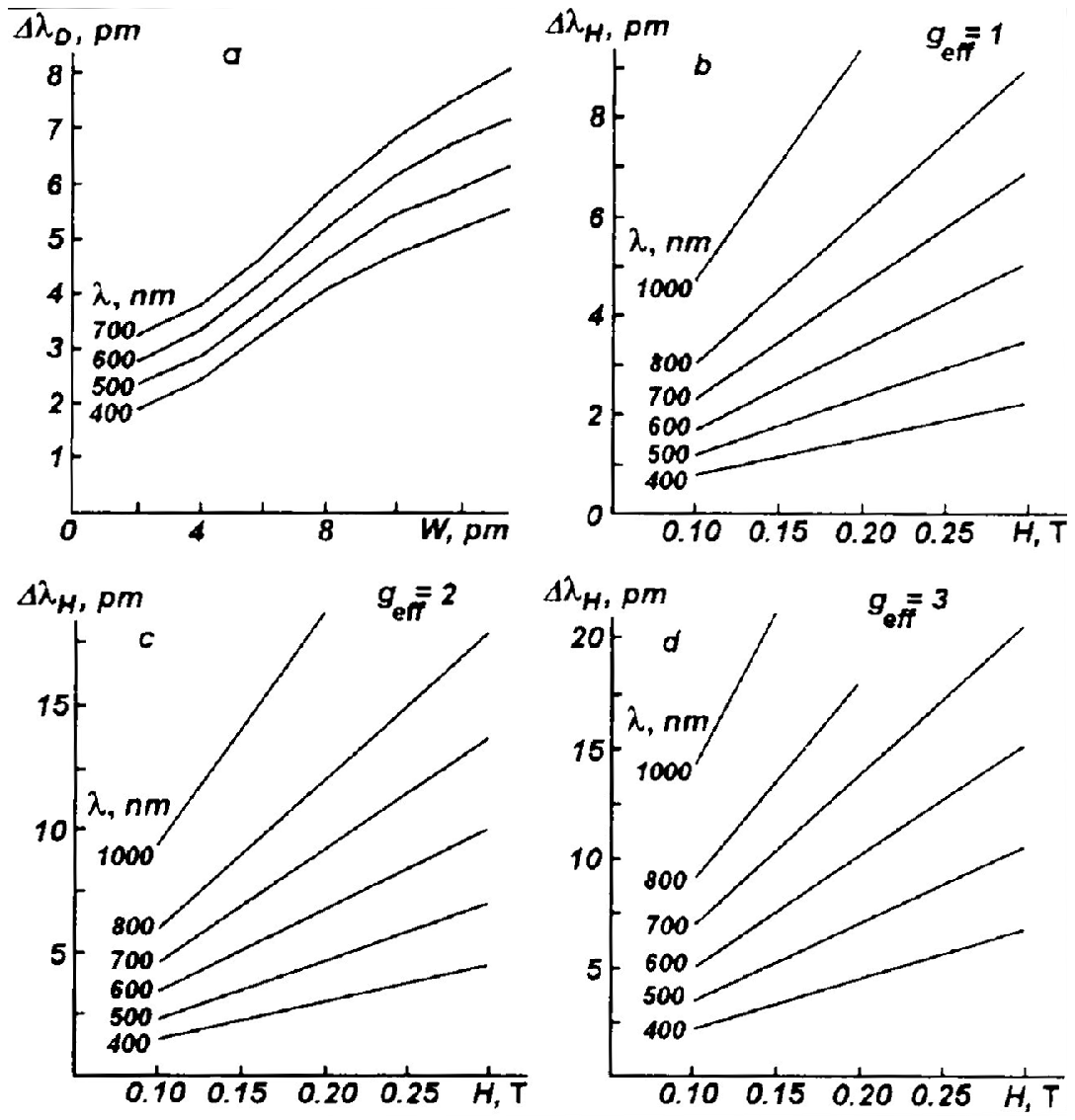}
\parbox[b]{15cm}{ \vspace{0.3cm}
{Fig. 8. Plots for the determination of magnetic field conditions for the Fe I
lines: a) half-width $\Delta\lambda_D$ v. equivalent width for lines not distorted
by magnetic field. Magnetic broadening $\Delta\lambda_H$ v. magnetic intensity $H$
for different wavelengths and effective Land\'{e} factors: b) $g_{\rm eff} = 1$, c)
$g_{\rm eff}= 2$, d) $g_{\rm eff}= 3$.

 } \label{Fig8}
 }

\end{figure}

\section{Conclusion}

Response functions are a fine means for the interpretation of observed Stokes
profiles. They allow a reliable determination of the depth of atmospheric layers
where any Stokes profile is disturbed by a disturbance in the temperature,
pressure, velocity, and magnetic field. We can find the sensitivity of the profiles
to fluctuations at any depth in the atmosphere, obtain quantitative sensitivity
estimates which simplify the analysis of profile sensitivity.

Being used in the spectral analysis, the sensitivity indicators, which are the net
response of all atmospheric layers where the line is formed to changes in one of
atmospheric parameters, allowed us to analyze quantitatively the sensitivity of
Stokes profiles by the example of iron lines. We made the following conclusions.

The sensitivity of the Stokes I profiles of photospheric lines to temperature,
pressure, and microturbulent velocity is generally the same as for the line
profiles in the absence of magnetic field \cite{3}. The magnetic sensitivity of the
Stokes V profiles is principally determined by the magnetic field conditions. The V
profiles are not sensitive to magnetic field variations under the strong-field
conditions ($\Delta\lambda_H \gg \Delta\lambda_D$) . They become sensitive under
the intermediate ($\Delta\lambda_H \approx \Delta\lambda_D$) and weak
($\Delta\lambda_H \ll \Delta\lambda_D$)  field conditions. Under specific field
conditions, the sensitivity of V profiles will be the strongest in lines with large
wavelengths, large effective Land\'{e} factors, large equivalent widths in media
with a high magnetic intensity. To make the temperature effect, which depresses the
magnetic sensitivity, less pronounced, we have to take into account the effective
excitation potential and to select lines with high excitation and ionization
potentials. Just these lines reveal an anomalous temperature sensitivity of their V
profiles --- the depression grows with $T$ rather than decreases. Chemical elements
like Si I, Fe II, O I, C I may have such lines.

Sensitivity indicator tables calculated for individual lines may be useful in
forming line pairs that are often used in the diagnostics of temperature and
magnetic stratification inside magnetic tubes. Sometimes it is necessary to select
a line with a predominant temperature or magnetic sensitivity. In this case it is
desirable to have tables of sensitivity indicators for every kind of atmospheric
parameters which affect the profiles and for a large number of spectral lines. Thus
it would be possible to find a line sensitive to one parameter and insensitive to
other parameters.


{\bf Acknowledgements.} We wish to thank A. S. Gadun for interest in this study and
S. K. Solanki for furnishing the DIAMAG program listing, which allowed an
additional test of the SPANSATM program, and for the opportunity to get acquainted
with the most complete overview \cite{11} of the latest studies on small-scale magnetic
fields.


\newpage

\end{document}